\documentclass[10pt]{article}
\newcommand{\be}{\begin{equation}}
\newcommand{\ee}{\end{equation}}
\newcommand{\ba}{\begin{eqnarray}}
\newcommand{\ea}{\end{eqnarray}}

\usepackage[english]{babel}
\usepackage{amsfonts,latexsym}
\usepackage{graphicx}
\usepackage{latexsym}
\usepackage[latin1]{inputenc}

\begin{document}

\title{Accurate light-time correction due to a gravitating mass}

\author{Neil Ashby \\ Department of Physics \\ University of Colorado, Boulder, Co. (USA) \\ \\
Bruno Bertotti \\ Dipartimento di Fisica Nucleare e Teorica \\ Università di
Pavia (Italy) }

\maketitle


\abstract{This technical paper of mathematical physics arose as an aftermath of Cassini's 2002
experiment \cite{bblipt03}, in which the PPN parameter $\gamma$ was measured with an
accuracy $\sigma_\gamma = 2.3\times 10^{-5}$ and found consistent with the prediction
$\gamma =1$ of general relativity. The Orbit Determination Program (ODP) of NASA's Jet
Propulsion Laboratory, which was used in the data analysis, is based on an expression
(\ref{eq:standardmoyer}) for the gravitational delay $\Delta t$ which differs from the
standard formula (\ref{eq:standard}); this difference is of second order in powers of $m$
-- the gravitational radius of the Sun -- but in Cassini's case it was much
larger than the expected order of magnitude $m^2/b$, where $b$ is the distance of closest
approach of the ray. Since the ODP does not take into account any other second-order
terms, it is necessary, also in view of future more accurate experiments, to revisit the
whole problem, to systematically evaluate higher order corrections and to determine which
terms, and why, are larger than the expected value. We note that light propagation in a
static spacetime is equivalent to a problem in ordinary geometrical optics; Fermat's
action functional at its minimum is just the light-time between the two end points A and
B. A new and powerful formulation is thus obtained. This method is closely connected with the much more general approach of \cite{poncinlafitte04}, which is based on Synge's world function.  Asymptotic power series are necessary
to provide a safe and automatic way of selecting which terms to keep at each order.
Higher order approximations to the required quantities, in particular the delay and the
deflection, are easily obtained. We also show that in a close superior conjunction, when
$b$ is much smaller than the distances of A and B from the Sun, of order $R$, say, the
second-order correction has an \emph{enhanced} part of order $m^2R/b^2$, which
corresponds just to the second-order terms introduced in the ODP. Gravitational
deflection of the image of a far away source when observed from a finite distance from the mass is obtained up to $O(m^2)$.}

\section{Introduction}

In the framework of metric theories of gravity and the PPN formalism, the
main violations of general relativity -- those linear in the masses -- are
described by a single dimensionless parameter $\gamma$. The question, at what
level and how general relativity is violated, in particular how much $\gamma$
differs from unity, Einstein's value, is still moot. No definite and consistent
prediction about it are available, except for the inequality $\gamma
<1$, which must be fulfilled in a scalar-tensor theory, in particular those
arising as the low-energy limit of certain string theories. To date, the best
measurement of $\gamma$ has been obtained with Cassini's experiment, which has
provided the fit (at 1-$\sigma$)

\be \gamma - 1 = (2.1 \pm 2.3)\times 10^{-5} \label{eq:cassini}.\ee Einstein's prediction
is still acceptable, but more accurate experiment are needed and planned.

While $\gamma$ controls also other relativistic effects, in particular those
related to gravito-magnetism, it mainly affects electromagnetic propagation.
The differential displacement of the stellar images near the Sun historically
was the first experimental effect to be investigated and is now of great
importance in accurate astrometry. The bending of a light ray also increases
the light-time between two points, an important effect usually named after its
discoverer I. I. Shapiro \cite{shapiro64}. Several experiments to measure this
delay have been successfully carried out, using \emph{wide-band} microwave
signals passing near the Sun and transponded back, either passively by planets,
or actively, by space probes (see \cite{will93}, \cite{reasenberg79}).

Cassini's 2002 experiment has implemented a third way to measure $\gamma$ \cite{bbgg92},
in which \emph{coherent} microwave trains sent from the ground station to the spacecraft
(at that time about 7 AU far away) were transponded back continuously. The use of
high-frequency carriers (in K$_a$ band, 34 and 32 GHz) and the combination with standard
X-band carriers (about 8 GHz) allowed successful elimination of the main hindrance,
dispersive effects due to the solar corona traversed by the beam. The tracking was
carried out around the 2002 superior conjunction; the minimum value of the impact
parameter of the beam was $1.6\, R_\odot$, but in effect only 18 passages have been used,
with a minimum impact parameter of $ \approx 6 \: R_\odot$. The two-way total amount of
phase between the time of emission and the time of arrival has been continuously measured
in each passage. In effect, however, NASA's Deep Space Network provides the phase count
in a given integration time $\tau$. Mathematically, in the limit $\tau \rightarrow 0$
this would give the received frequency, in which Doppler effects and gravitational
frequency shift are mixed up (Sec. 4). Cassini's observable, therefore, can also be
assessed in terms of the predicted change in frequency, as in \cite{bbgg92}; but in
practice, taking $\tau$ small would introduce unacceptable high-frequency noise. The
change in light-time in a given integration time is the correct, theoretically available
observable.

In the standard formulation for a superior conjunction, and taking the Sun at rest, the (one-way) light-time from
an event A to an event B is:

\be t_{_B} - t_{_A} = r_{_{AB}} + \Delta t = r_{_{AB}} +(1+ \gamma)m \ln \frac{r_{_A}+
r_{_B} + r_{_{AB}}}{r_{_A}+ r_{_B} - r_{_{AB}}}, \label{eq:standard}\ee where $m = 1.43 $
km is the gravitational radius of the Sun, $r_{_A}, r_{_B}$ are, in Euclidian geometry
(See Fig. 1 left), the distances of A and B from the Sun and $r_{_{AB}}$ their distance.
The velocity of light $c$ is unity. $\Delta t$, the increase of the light-time over
$r_{_{AB}}$, is the \emph{gravitational delay}.

In a \emph{close superior conjunction} A and B are on the opposite sides of the mass and
the Euclidian distance $b_0$ of the straight line AB from the mass fulfils $b_0 \ll
(r_{_A},r_{_B}) =O(R)$, say. In this approximation eq. (\ref{eq:standard}) reduces to

\be t_{_B} - t_{_A} = r_{_{AB}} + \Delta t = r_{_{AB}} +(1+ \gamma)m\ln
\left(\frac{4r_{_A} r_{_B}}{b_0^2}\right), \label{eq:conjunction} \ee with a logarithmic
enhancement over the formal order of magnitude $\Delta t = O(m)$.\footnote{As stated in
the supplementary material, in eq. (2) of \cite{bblipt03} the two terms in the right-hand
side should obviously be multiplied by a factor 2. This error, of course, had no
consequence on the computer fit.} Taking the logarithm equal to 10, this provides an
estimate of the timing accuracy in terms of the error in $\gamma$:

\be \sigma_{_{\Delta t}} = 1.43 \,\sigma_\gamma\times 10^6 \,\mathrm{cm},
\label{eq:sensitivity}\ee  corresponding, in Cassini's case, to 30 cm.
(\ref{eq:conjunction}) embodies also the one-way frequency change $\Delta \nu$ induced by
gravity between A and B. Their motion makes $b_0$ (and the distances) change with time,
so that, for a one-way experiment,

\be \frac{\Delta \nu}{\nu} = \frac{d \Delta t}{dt} = -2(\gamma +1)
\frac{m}{b_0}\frac{db_0}{dt}. \label{eq:frequency}\ee

The basic geometric setup is straightforward: a point mass $m$ at rest at the origin in
an asymptotically flat space generates a line element with rotational symmetry. An
invariant Killing time $t$ is defined; events on each $t = $ constant surface are
`simultaneous' and the metric components are constant. The proper time $ds =
\sqrt{g_{00}(r)}\,dt$ of a static observer differs from $dt$ by the red-shift factor
$\sqrt{g_{00}(r)}$. A null geodesic runs from the event A (with radial coordinate
$r_{_A}$ and time $t_{_A}$) to the event B (with radial coordinate $r_{_B}$ and time
$t_{_B}$); it stays on a plane, taken here as the equatorial plane $\theta = \pi/2$. The
(invariant) longitude difference $\Phi_{_{AB}} =\phi_{_B}- \phi_{_A}$ completes the
setup. In the PPN formalism and isotropic coordinates the metric reads:

\ba ds^2 &=& A(r) dt^2 - B(r) d\ell^2 = \nonumber \\
&=& \left(1-\frac{2m}{r}+2\beta\frac{ m^2}{r^2}+ \ldots \right)dt^2- \left(1 +
\gamma\frac{2m}{r}+ \frac{3 \epsilon}{2}\frac{m^2}{r^2} + \ldots\right)d\ell^2
\label{eq:PPN},\ea where

$$d\ell^2 = dr^2 + r^2(d\theta^2 + \sin^2 \theta\, d\phi^2) = dr^2 + r^2 d\Omega^2 $$
is the Euclidian line element. The parameters $\gamma$, $\beta$ and $\epsilon$ are equal
to 1 in general relativity; while $\gamma$ and $\beta$ are accurately known, currently no information is available about $\epsilon$.

In our case the best mathematical tool to deal with electromagnetic propagation is not
null geodesics, but the theory of eikonal. It is known (e.g., \cite{misner73}) that in
this problem Fermat's Principle holds, corresponding to the refractive index

\be N(r) = \sqrt{\frac{B(r)}{A(r)}}; \label{eq:index}\ee we develop \emph{ab
initio} the eikonal and solve for it by separation of variables (Sec. 4). The
radial part provides Fermat's action as a radial integral containing $N(r)$ and
the impact parameter $h$; when computed at the true value $h_{\mathrm{true}}$,
such action is just the required light-time. The solution can be obtained
recursively, using appropriate expansions in powers of $m$: the expansion for
$h$ begins with $h_0 =b_0$, the distance of the straight line AB from the
origin. In this way the variational nature of the problem brings about a great
conceptual and algebraic simplification. At the linear approximation in $m$ one
would expect that the light-time contains $h_1$, the correction in the impact
parameter linear in the mass; as one can see from (\ref{eq:standard}), this is
not the case. This property is generally true: the correction to the light-time
$O(m)^k$ \emph{does not contain} $h_k$ (Sec. 6).

Cassini's and many other space experiments have been analyzed using NASA's Orbit
Determination Program (ODP), developed by NASA at Jet Propulsion Laboratory in the 60's
and steadily improved since; a new version called MONTE is under development. The ODP,
whose theoretical formulation is due to T. D. Moyer \cite{moyer00}, integrates the
equations of motion of the relevant bodies and provides their trajectories in the
\emph{ephemeris time}. This task is carried out in a reference system -- called BCRS
(Barycentric Coordinate Reference System) -- in which the centre of gravity of the solar
system is at rest and the Sun moves around with a velocity $v_\odot \approx 10
\,\mathrm{m/sec} = 3 \times 10^{-8} c$. As discussed in \cite{bbnali07}, the light-time
in this this frame differs from the rest frame of the Sun essentially due to Lorentz time
dilatation; being of order $v_\odot$, this difference is quite below the sensitivity of
Cassini's experiment. We do not discuss this point any more; $t$ is just Killing time.

The ODP uses a fictitious Euclidian space $S_3(x,y,z)$, which corresponds to the
isotropic coordinates of (\ref{eq:PPN}). This space is just a computational convenience
and should not be considered as a physical background in which gravity acts. For example,
replacing $r$, the Euclidian distance from the origin, with $r + km$, where $k$ is an
arbitrary constant, is fully legitimate in a covariant theory, but it destroys the
conformal flatness of space, introduces a gravitational potential $-km^2/r^2$ and adds a
second-order term to the delay $\Delta t$. Strictly speaking, the word `delay' is
inappropriate: we just have a light-time and there is nothing with respect to which a
delay can be reckoned. The object of the measurement is the time change of the delay.  The arbitrariness of the radial coordinate also affects
gravitational bending: its second-order approximation up to $O(m/b)^2$, depends on which
radial coordinate is used (see \cite{fischbach80}, \cite{epstein80} and \cite{richter82})
\cite{bodenner03}).

It should also be noted that the spacetime coordinates of the end events are not directly
provided in the experimental setup and depend on the gravitational delay $\Delta t$, the
very quantity one sets out to measure. The trajectories $\mathbf{r}_{_{A}}(t)$ and
$\mathbf{r}_{_{B}}(t)$ are given by the numerical code; the starting time $t_{_{A}}$ is
just a label of the ray, but the arrival time $t_{_{B}}$ is greater than $t_{_{A}} +
r_{_{AB}}$. The way out is to take for the end point

$$ \mathbf{r}_{_{B}}(\mathbf{r}_{_{B}}) = \mathbf{r}_{_{B}}(t_{_{A}} +r_{_{AB}}) + \Delta
t\, \mathbf{u}_{_{B}}(t_{_{A}}+ r_{_{AB}}), $$ where $\mathbf{u}_{_{B}} =
d\mathbf{r}_{_{B}}/dt$ For a typical velocity $10^{-4}\,c$ the correction is of order $20
\times 1.4 \times 10^5 \times 10^{-4} = 300$ cm, and the \emph{a priori } accuracy in
$\Delta t$ is sufficient.

Since for electromagnetic propagation $dt$ and $d \ell$ in (\ref{eq:PPN}) are almost
equal, (\ref{eq:standard}) is the correct approximation to the delay to $O(m)$; one would
expect this to be the first term in an expansion in powers of $m/b_0$, so that the next
term should be

$$  \approx m\frac{m}{b_0} = m\frac{m}{R_\odot} \frac{R_\odot}{b_0}
=0.3\,\frac{R_\odot}{b_0}\, \mathrm{cm}, $$ quite below Cassini's sensitivity. The
present paper arose because the ODP (eq. (8-54) of \cite{moyer00}), in fact does not use
(\ref{eq:standard}), but, in our notation,

\be t_{_B} - t_{_A} = r_{_{AB}} + \Delta t = r_{_{AB}} +(1+ \gamma)m
\ln\left(\frac{r_{_A}+ r_{_B} + r_{_{AB}} + (1+ \gamma)m}{r_{_A}+ r_{_B} - r_{_{AB}}+ (1
+ \gamma)m}\right).   \label{eq:standardmoyer}\ee We have not been able to fully
reconstruct Moyer's derivation of this expression. It introduces non linear corrections
arising from non linear effects of linear metric terms, but no quadratic metric terms. However,
the difference between the two expressions of the delay is much larger than the estimate
above; this arises because in Cassini's case, in (\ref{eq:standard}) the denominator
$r_{_{A}} + r_{_{B}} -r_{_{AB}}$ is much smaller than the numerator $\approx 2r_{_{AB}}$.
Indeed,

\be \Delta t - (\Delta t)_{_{\mathrm{ODP}}} = -2(1+\gamma)^2
\frac{m^2}{b_0^2}\frac{r_{_{A}}r_{_{B}}}{r_{_{A}} + r_{_{B}}} = -(1+\gamma)^2\frac{m^2
R}{b_0^2} ,\label{eq:moyercorrection}\ee where we have introduced the harmonic mean of
the distances

\be \frac{2}{R} = \frac{1}{r_{_{A}}} + \frac{1}{r_{_{B}}}= \frac{r_{_{A}} +
r_{_{B}}}{r_{_{A}}r_{_{B}}} \label{eq:harmonic}. \ee If, as in Cassini's experiment,
$r_{_{B}} \gg r_{_{A}} = 1\, \mathrm{AU} = 200 \,R_\odot$, $ R = 400 \,R_\odot$ the
correction is about

$$ 1600\, m \frac{m}{R_\odot} \left(\frac{R_\odot}{b_0} \right)^2 =
500 \left(\frac{R_\odot}{b_0} \right)^2 \mathrm{cm } .$$ Even at $\approx 6\, R_\odot $
this correction is somewhat below the sensitivity (\ref{eq:sensitivity}) and it should
not have affected the result. However, it cannot be excluded that
neglected non linear terms relevant for Cassini's experiment affect the fit
(\ref{eq:cassini}).  One could say, (\ref{eq:standardmoyer}) is mendacious; a full clarification of the
problem is needed.

Empirically dropping or keeping `small' terms may lead to inconsistencies and does not
work; the rigourous method of \emph{asymptotic perturbation theory} (see, e. g.,
\cite{erdely56}, \cite{hinch91}) must be used. We briefly sketch it now at a practical
level. One begins with a wise choice of a dimensionless `smallness' parameter, and
expands every function in the corresponding power series. Our main choice will be
$m/b_0$, but convenience may suggest using other lengths, like in $m/r$. An asymptotic
series

$$ G = \sum_sG_s \left(\frac{m}{b_0}\right)^s $$ is a formal object assigned just by the sequence of its
coefficients $G_s$; arithmetics and calculus follows the obvious rules for sum,
multiplication and differentiation.  Equality between two asymptotic series just means
that the coefficients of the same order are equal. The value of $G(m)$ as a function of
$m$ does not play any role, and even the convergence of the series is irrelevant; what
matters is only the truncated value at any order $k$

\be  G_{(k)} = \sum_{s=0}^k\left(\frac{m}{b_0}\right)^s G_s +
O\left(\frac{m}{b_0}\right)^{k+1}. \ee The parameter should not be understood as a fixed
number, but as a variable which tends to zero. The symbol $O(.)$ means \emph{ order of
infinitesimal}; it states how fast the remainder tends to zero as the parameter
diminishes. An asymptotic series can be constructed from an ordinary arbitrary function
$G(m)$; but a whole class of functions give rise to the same series; for example, if
$G_s$ is the sequence generated by $G(m)$, the same sequence is also generated by

$$ \big[1 +P\exp(-Qb_0/m)\big]G(m) \quad (Q > 0). $$ In this way any recursive iteration then proceeds
automatically and safely, even in the most complex situations.

In our case light-time will be provided as an asymptotic power series

\be t_{_{B}} - t_{_{A}} = r_{_{AB}} + m\sum_{s=1}
\Delta_s\left(\frac{r_{_{A}}}{b_0},
\frac{r_{_{B}}}{b_0}\right)\left(\frac{m}{b_0}\right)^{s-1},
\label{eq:delayDelta} \ee with dimensionless coefficients $\Delta_s$ .
$\Delta_1$ provides the lowest, standard approximation to $\Delta t$ (see
(\ref{eq:standard}). In principle, asymptotic analysis does not provide a
numerical estimate of the remainder in a given situation; this is a physical,
not a mathematical question. But when the problem, properly formulated, does
not contain small dimensionless quantities other than the smallness parameter
itself, one can expect the mathematical operations leading to the result to
maintain the order of magnitude and to lead to expansions whose coefficients
are numerically of the same order. This is the case of deflection, the angle
between the asymptotes of the ray. There is only one length in the problem, the
distance $b$ of the point of closest approach, or, equivalently, the impact
parameter $h = bN(b)$ (see Fig. 4); hence in the expansion

\be \delta = \sum_s \delta_s \left(\frac{m}{h}\right)^s
\label{eq:deflection}\ee the coefficients $\delta_s$ are dimensionless numbers,
solely determined by the PPN parameters and, must be of order unity (see Sec.
9). But in the delay problem the coefficients $\Delta_s$ depend on the
geometrical configuration. They are of order unity in the generic (but scarcely
interesting) case in which $ r_{_A}, r_{_B}$ and $b_0$ are of the same order;
but in a close superior conjunction  -- of crucial relevance in experimental
gravitation -- when $b_0 \ll (r_{_{A}}, r_{_{B}}) =O(R)$, besides $m/b_0$,
there is another smallness parameter, namely, $b_0/R$, and there is no reason to exclude
that the $\Delta_s$ increase with $R/b_0$ beyond the expected order of
magnitude unity. This we call \emph{enhancement}.  We already saw in (\ref{eq:conjunction}) that $\Delta_1$ is enhanced, albeit only logarithmically; the ODP correction (\ref{eq:moyercorrection}), formally of second order, is enhanced by $R/b_0$. This could place serious limitations on the method and even invalidate the iteration itself. This would occur, for instance, when $mR \approx b_0^2$; if $b_0 = R_\odot = 1/200 AU$, this corresponds to $R = 2000$ AU.  The enhancement, which has never been discussed in the literature, has been fully understood and tamed in the present paper (Sec. 8). We have found, indeed, that \emph{the second-order terms embodied in the ODP expression (\ref{eq:standardmoyer}) which was used in Cassini's experiment are just the enhanced second-order terms;} Cassini's result (\ref{eq:cassini}) is still safe.

The problem can be reduced to one of ordinary optics; due to its variational nature, the eikonal function can be easily solved in an expansion in powers of $m/h$.  The second-order expression of the light-time for a static spacetime has been obtained; extension to third order is also easy.  This approach should be compared with the much more general work of \cite{poncinlafitte04}, who consider Synge's world function $\Omega(x_{_A},x_{_B})$ in a generic spacetime for a generic geodesic (not necessarily null) between two events A and B.  On the basis of Hamiltonian theory, they develop a method to solve for $\Omega(x_{_A},x_{_B})$ in a formal power series with respect to the gravitational constant $G$ and compute it up to the second order.  In the null case the world function vanishes on the solution and becomes the eikonal function.  Out method, limited of course to the spherically symmetric case, exploits directly the variational nature of the problem and leads to the second-order expression of the light-time, which agrees with the expression of \cite{poncinlafitte04}; extension to third order is also easy.  

For a realistic observation of a distant source from a point B at a finite distance $r_{_{B}}$,
(\ref{eq:deflection}) must be generalized to an expansion of the type
\be \delta_{_{B}} = \sum_s \delta_{Bs}
\left(\frac{r_{_{B}}}{h}\right)\left(\frac{m}{h}\right)^s
\label{eq:deflectionB},\ee where $h$ is the impact parameter. The linear term
has been evaluated in \cite{misner73}, \$ 40.3; the quadratic correction will
be obtained in Sec.9 .

\section{Hyperbolic Newtonian dynamics}

Newtonian dynamics of a test particle attracted by a point mass $M$, an exactly
soluble problem, illustrates these issues. We consider a motion in the
equatorial plane $\theta = \pi/2$, with radial coordinate $r$ and azimuthal
longitude $\phi$. The Lagrangian function

\be \mathcal{L}_{\mathrm{New}} = \frac{1}{2}\left[\left(\frac{dr}{dt}\right)^2
+ r^2\left(\frac{d\phi}{dt}\right)^2\right] + \frac{GM}{r} \ee keeps the total
energy $v_\infty^2/2$ constant; $v_\infty$, the ultimate speed of the particle
at a large distance, plays a role analogous to the speed of light and will be
taken equal to unity. Then

\be \left(\frac{dr}{dt}\right)^2 + r^2\left(\frac{d\phi}{dt}\right)^2
-2\frac{m}{r} = 1, \ee where $m = GM/v_{\infty}^2$ is the gravitational radius.
$\phi$ is an ignorable coordinate, so that the angular momentum

\be \label{angmom} \frac{\partial \mathcal{L}_{\mathrm{New}}}{\partial
(d\phi/dt)} = r^2 \frac{d\phi}{dt}=h \ee is constant. Since the velocity at
infinity is 1, $h$ is also the impact parameter. Eliminating $dt$ we get:

\be r\frac{d\phi}{dr}= \pm \frac{h }{\sqrt{r(r+2m )-h^2}}\,; \label{eq:angmom}\ee hence

\be h=\sqrt{b(b+2m )} \ee determines $b$, the distance of closest approach
where $dr/d\phi =0$. The sign depends upon whether the ray is ingoing or
outgoing. Integrating we get the true anomaly

\be f=\arccos \left(\frac{b^2+2 m b-m  r }{r(b+m )} \right)\,. \ee Alternatively,
the motion can be expressed in terms of the semi-major axis $a = m $ and the hyperbolic eccentricity
 $e = 1+b/m $:

\be r=\frac{a(e^2-1)}{1+e \cos f}\,. \ee The acute angle $\delta$ between the
asymptotes is given by

\be \sin \delta = \sin\left(2 \arccos \left(-\frac{1}{e} \right)\right) =\frac{2m }{b+m
}\sqrt{1-\frac{m^2}{(b+m )^2}} \ee This angle has a regular expansion in powers of $m/b$,
with no enhancement.

\begin{figure}
\includegraphics[width=4in]{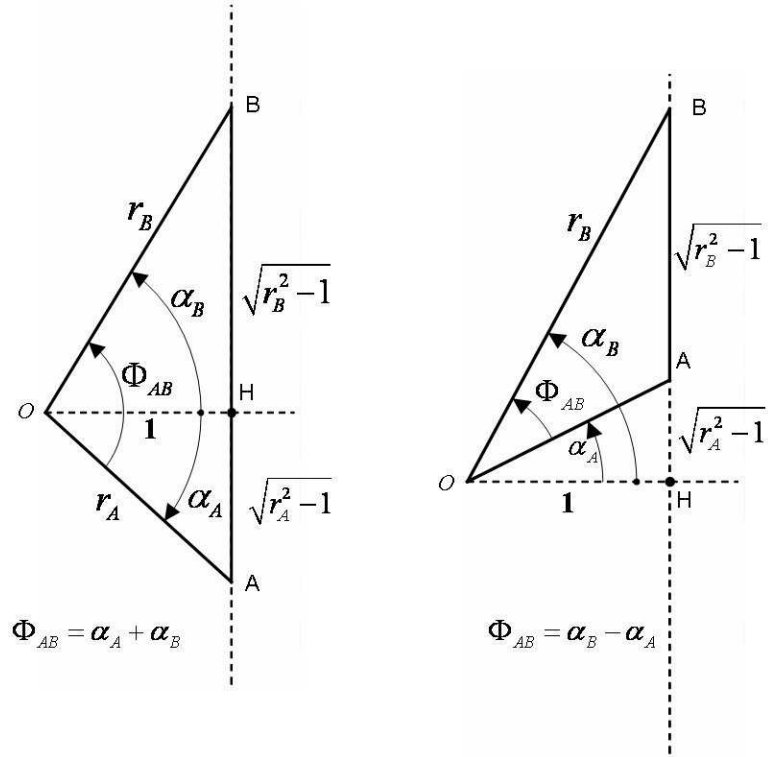}
\caption{The background Euclidian geometry. The mass at O and the end points at
A and B define a triangle AOB; the distance  $OH = b_0$ from the straight line
AB to the mass at the origin is taken as the unit of length. The angles
$\alpha$ are taken positive. The internal angle $\Phi_{_{AB}}$ can be obtuse
(left) or acute (right); in the first, more interesting case, when, in
addition, $r_{_{B}} \geq r_{_{A}} \gg b_0$, we have the most important case of
a close superior conjunction, in which the deflection is large. Elementary
trigonometry gives the relation $ b_0\sqrt{r_{_{A}}^2 + r_{_{B}}^2 - 2r_{_{A}}
r_{_{B}}\cos \Phi_{_{AB}}} = r_{_{A}} r_{_{B}} \sin  \Phi_{_{AB}}$\,.}
\end{figure}

Consider, however, the hyperbola determined by two points $A$ and $B$ on the opposite
sides of the vertex (right side in Fig. 1).  As in space navigation -- in
particular in the ODP -- the end points are provided in terms of the initial and final
position vectors, or equivalently, in terms of the initial and final distances $r_{_{A}}$
and $r_{_{B}}$ and the elongation angle $\Phi_{_{AB}}$; the ``unperturbed distance of
closest approach" may then be calculated from elementary geometry:

\be b_0= \frac{r_{_{A}}r_{_B} \sin \Phi_{_{AB}}}{\sqrt{r_{_A}^2+r_{_B}^2-2 r_{_A} r_{_B}
\cos \Phi_{_{AB}}}}\,. \ee Choosing the angles $\alpha_{_A}$, $\alpha_{_B}$ and
$\Phi_{_{AB}}$ positive, we can express $b$ in terms of $b_0$ with the condition

\ba \Phi_{_{AB} } &=& \alpha_{_A}+\alpha_{_B} = \arccos\frac{b_0}{r_{_A}} +
\arccos\frac{b_0}{r_{_B}}  = \nonumber \\
& = &\arccos\left(\frac{b(b+2m )}{r_{_{A}}(b+m )}- \frac{m }{b+m }\right) + \nonumber \\
&+ &\arccos\left(\frac{b(b+2m )}{r_{_{B}}(b+m )}-\frac{m }{b+m }\right)\,. \ea The symmetric case
$r_{_{A}}= r_{_{B}}=R$ is sufficient to exhibit the problem. The condition reads:

\be  \frac{b}{R} \frac{b+ 2m }{b +m } -\frac{m }{b +m } - \frac{b_0}{R} =0\,, \ee or \be
b^2 + (2m  -1)b - m (R+ 1) =0\,, \ee with the solution

\be \frac{2b}{b_0} = \frac{b_0 -m}{b_0} + \sqrt{1 + 4\frac{m}{b_0} \frac{R +
m}{b_0}}. \ee Expansion in powers of $m $ gives

\be \frac{b}{b_0} = 1 + \left(\frac{R}{b_0}-1\right)\frac{m}{b_0}  -
\left(\left(\frac{R}{b_0}\right)^2-1\right)\left(\frac{m}{b_0}\right)^2 +
O\left(\frac{mR}{b_0}\right)^3. \label{eq:b_exp} \ee The enhancement is clear:
when $R =O(b_0)$ the truncation error at order $k$ is $O(m/b_0)^{k+1}$, with a
coefficient of order unity, as na\"{\i}vely expected; but when -- as in a close
superior conjunction -- $R \gg b_0$, the error is larger, $O(mR/b_0^2)^{k+1}$.
Formally this requires introducing another smallness parameter $b_0/R$ and
expanding every coefficient of the primary $m$-expansion in descending powers
of $R/b_0$.  Of course, the condition
\be \frac{mR}{b_0^2} \ll 1, \label{eq:quantity} \ee must be fulfilled, lest the whole procedure breaks down.  One could say, anchoring the trajectory at far away end points has a lever effect, so that an increase in the mass produces a large increase in closest approach.  

The quantity (\ref{eq:quantity}) gives, in order of magnitude, the ratio between the deflection $\approx m/b_0$ and the angle $b_0/R$ which separates the central mass and a distant star, as seen from a distance $R$. Hence the limiting constraint above implies that the geometry of astronomical deflection is the same as in the classical case (see Fig. 4): sources in the sky near the Sun are displaced outward by an amount inversely proportional to the angular distance.  The transition through the milestone $mR = b_0^2$ marks the passage to the gravitational lensing regime, in which the image can appear on both sides.

In Sec. 8 the light-time enhancement is dealt with in the general case
and it is shown that the dimensionless coefficients $\Delta_s$ in
(\ref{eq:delayDelta}) are $O(R/b_0)^{s-1}. $

\section{The radial gauge}

The metric of a spherical body at rest has the general form

\be ds^2 = A(r) dt^2 - B(r) dr^2 - C(r)r^2 d\Omega^2,\ee where $A(r), B(r),
C(r)$ are of power series the form:

\be A(r) = \sum_s A_s\left(\frac{m}{r}\right)^s. \ee It is asymptotically flat, so that
$A_0 = B_0 = C_0 = 1 $. The radial coordinate is otherwise arbitrary; this is the
\emph{gauge freedom} at our disposal. For consistency, however, any change $r \rightarrow \bar{r}
= g(r)$ must become an identity at infinity and have a similar expansion:

\be g(r) = r + g_1m + g_2 \frac{m^2}{r} + \ldots ;\ee the coefficients $A_s, B_s, C_s$
are not gauge invariant. Two gauges are common. In the \emph{isotropic} form -- the
canonical choice in space physics -- $C(r)=B(r)$, so that

\be ds^2 = A(r) dt^2 - B(r)(dr^2 + r^2 d\Omega^2) =A(r) dt^2 - B(r)d\ell^2;
\label{eq:isotropic} \ee the space part of the metric is conformally flat. We define

\be N(r) = \sqrt{\frac{B(r)}{A(r)}} = \sum_s N_s \left(\frac{m}{r}\right)^s = 1
+N_1\frac{m}{r} + N_2\left(\frac{m}{r}\right)^2 +O\left(\frac{m}{r}\right)^3.
\label{eq:linear}\ee  In the PPN scheme (e. g., \cite{will93})

\be N_1 = \gamma+1, \quad N_2 = \frac{6 -4\beta + 3\epsilon + 4\gamma
-2\gamma^2}{4}. \label{eq:N2}\ee In `Schwarzschild' gauge $\bar{C}(\bar{r})=1$
and

$$ ds^2 = {\bar A}(\bar{r}) dt^2 - {\bar B}(\bar{r})d\bar{r}^2 - \bar{r}^2
d\Omega^2; $$ the area of a sphere of radius $\bar{r}$ is just the Euclidian
expression $4\pi \bar{r}^2$, which defines $\bar{r}$ in an invariant way. In
the original Schwarzschild solution ${\bar A}(\bar{r})= 1/\bar{B}(\bar{r})= 1 -
2m\gamma/\bar{r}$. To get the isotropic form one requires

\be g^2(r) = {\bar B}(g(r))\left( \frac{d g}{dr}\right)^2 r^2; \ee to first
order

\be \bar{r} = r + \gamma m + \ldots\, \label{eq:gauge}. \ee In the present paper a third radial coordinate

\be \rho = r N(r) = r\sqrt{\frac{B(r)}{A(r)}} = r + mN_1 + m^2 \frac{N_2}{r} + \ldots
\label{eq:moyer} \ee plays an important role. It is a monotonic function of $r$ and
ensures $A(\rho) =C(\rho)$. In the linear approximation it was introduced by Moyer in
\cite{moyer00} (eq. (8-23)), and boils down to just adding to $r$ a constant term, equal
to 2.95 km for the Sun.

\section{Geometrical optics}

It is convenient to reduce the problem to geometrical optics using the eikonal
function $\mathfrak{S}$. In a generic spacetime $\mathfrak{S}$ fulfils the
eikonal equation

\be g^{\mu \nu} \partial_\mu \mathfrak{S}\,\partial_\nu \mathfrak{S} = 0;  \ee
its characteristics are the null rays (see, e. g., \cite{bel94}).
$\mathfrak{S}$ is the phase of the electromagnetic wave.  Let $r_{_{A}}^\mu=r^{\mu}(s_{_{A}}), r_{_{B}}^\mu =r^{\mu}(s_{_{B}})$ be the trajectories of the end points, given as functions of their proper times $s_{_A}, s_{_B}$; let
$$ v_{_{A}}^\mu = \frac{d r^\mu}{ds_{_{A}}}, \quad v_{_{B}}^\mu = \frac{d
r^\mu}{ds_{_B}} $$ be the corresponding four-velocities. Clocks associated with
them measure the proper frequencies

\be \omega_{_{A}} = -v_{_{A}}^\mu \partial_\mu \mathfrak{S} = \frac{d
\mathfrak{S}}{ds_{_A}}, \quad \omega_{_{B}} = -v_{_{B}}^\mu \partial_\mu
\mathfrak{S} = \frac{d \mathfrak{S}}{ds_{_B}}. \ee In the simple case in which
the end points are far away from the source, where the metric corrections can be neglected, the
contribution to the frequency difference corresponds to the ordinary Doppler
effect, and can be evaluated with a slow motion expansion; the change in
$\mathfrak{S}$ between A and B is determined by the accumulated gravitational
effect along the ray and mainly come from the region near the mass. 
\be g^{\mu \nu} \partial_\mu \mathfrak{S}\,\partial_\nu \mathfrak{S} = 0 =
N^2(\mathbf{r})(\partial_t\mathfrak{S})^2 -  \nabla \mathfrak{S}  \cdot \nabla
\mathfrak{S}, \label{eq:hamilton} \ee where $\nabla$ is the Euclidian gradient
operator. We are really interested only in the spherically symmetric case, but
the reasoning of this Section holds also for an arbitrary $N(\mathbf{r})$.

$\mathfrak{S}$ is the phase; propagation occurs keeping it constant. Separating
space and time variables with

$$ \mathfrak{S}= \mathfrak{S}_t(t) + \overline{\mathfrak{S}}_{\bf r}(\mathbf{r}), $$
leads to the class of solutions

\be \mathfrak{S} = \omega_0\big(\overline{\mathfrak{S}}(\mathbf{r}) -t\big),
\ee where $\omega_0 \overline{\mathfrak{S}}(\mathbf{r})$ is the spatial part of
the phase and $\omega_0$ is a constant frequency. $\overline{\mathfrak{S}}$ has
the dimension of time and satisfies

\be \nabla \overline{\mathfrak{S}} \cdot \nabla \overline{\mathfrak{S}} =
N^2(\mathbf{r}) \label{eq:eikonal}. \ee If a clock is at rest relative to the
mass, $v^\mu = (1, \mathbf{0})/\sqrt{A(r)}$, and the measured proper frequency
$\omega_0/\sqrt{A(r)}$ includes the appropriate gravitational shift away from
the asymptotic value $\omega_0$. This is enough to reduce the problem to
geometrical optics (see, e. g., \cite{born64}, Ch. III). A ray
$\mathbf{r}(\ell)$, as function of the Euclidian arc length $\ell$, is
orthogonal to the eikonal surfaces $\overline{\mathfrak{S}}(\mathbf{r}) =
\mathrm{const}$ and fulfils

\be \frac{d}{d\ell}\left(N(\mathbf{r}) \frac{d\mathbf{r}}{d\ell }\right) =
\nabla N(\mathbf{r}). \label{eq:trajectory}\ee The index of refraction is the
rate of increase of the spatial phase along the ray:

$$ \frac{d\overline{\mathfrak{S}}}{d\ell} = N(\mathbf{r}). $$
Consider now Fermat's action functional

\be S[\mathbf{r}(\lambda)]= \int_{\lambda_A}^{\lambda_B} d\lambda
N(\mathbf{r})\sqrt{\frac{d\mathbf{r}}{d\lambda}\cdot\frac{d\mathbf{r}}{d\lambda}}
= \int_{\lambda_A}^{\lambda_B} d\lambda\, \mathcal{L}_{_{F}}, \label{eq:FA}\ee
where the trajectory, any path joining the end points, is expressed in terms of
a generic parameter $\lambda$:

\be \mathbf{r}(\lambda_{_{A}})= \mathbf{r}_{_{A}}, \quad
\mathbf{r}(\lambda_{_{B}})
 = \mathbf{r}_{_{B}}. \label{eq:boundary} \ee Since the action is, in fact, independent
of the choice of $\lambda$, no generality is lost if $d\lambda = d\ell$, the
Euclidean line element. The Euler-Lagrange equation for the action
(\ref{eq:FA}) reduces to (\ref{eq:trajectory}). The actual elapsed time

\be t_{_{B}} - t_{_{A}} =  S(A,B) = \int_{\ell_A}^{\ell_B} d\ell N(r) =
\overline{\mathfrak{S}}_{_{B}} - \overline{\mathfrak{S}}_{_{A}}
\label{eq:action}\ee is just the value of $S[.]$ computed at a local minimum --
the actual ray (\emph{Fermat's Principle}). One should keep in mind the
distinction between the action functional, with its argument in square
brackets, and the action computed at the extremum, an ordinary function of the
end points denoted with $S(A,B)$. In $S(A,B)$, but not in $S[.]$, it is allowed
to replace the generic independent variable $\lambda$ with a more convenient
one related to the solution, like $r$. For simplicity, the different functions
denoted by the symbol $S$ are distinguished by their arguments; below, the
quantity $S(r_{_{A}}, r_{_{B}}; b) = S(h)$ will be introduced to denote the
action corresponding to a ray anchored at $r_{_{A}}$ and $r_{_{B}}$, but with
arbitrary $b$ (or $h$).
                                                      
\section{The solution}

The eikonal function provides a deep simplification in the evaluation of the
light-time. Having already separated out the time, the three-dimensional
eikonal equation (\ref{eq:eikonal}) in spherical symmetry and in the equatorial
plane can be solved by separating out the longitude $\phi$: setting

$$ \overline{\mathfrak{S}}(r,\phi) = \overline{\mathfrak{S}}_r(r) +
\overline{\mathfrak{S}}_\phi(\phi). $$ It satisfies \footnote{For a function of a single
variable a prime indicates the derivative.}

$$ r^2\big(\overline{\mathfrak{S}}_r'\big)^2 + \big(\overline{\mathfrak{S}}_\phi'\big)^2 =
r^2N^2(r), $$ so that $\overline{\mathfrak{S}}_\phi'$ is a constant. Setting $
\overline{\mathfrak{S}}_\phi =  h \phi$, the eikonal equation reduces to

$$ \big(\overline{\mathfrak{S}}_r'\big)^2 = \frac{1}{r^2}(r^2 N^2(r) - h^2),$$ with the primitive

$$ \overline{\mathfrak{S}}(r) = \pm \int^r \frac{dr}{r}\sqrt{r^2N^2(r) - h^2}. $$
The $+$ and the $-$ signs correspond, respectively, to an outgoing and an incoming
photon. The radial coordinate of closest approach $b$, where $\overline{\mathfrak{S}}_r'
=0$, is the solution of

\be bN(b) = h ;\label{eq:closest}\ee since $r \geq b$, $\mathfrak{S}$ is a real function.
In Sec. 9 it will be shown that $h$, just like in the Newtonian case, is the impact
parameter (Fig. 4). The total phase is, therefore,

\be \mathfrak{S}= \omega_0\left( h\phi \pm  \int^r \frac{dr}{r}\sqrt{r^2N^2(r) - h^2} -
t\right).\ee A wavefront propagates keeping $\mathfrak{S}$ constant, so that the time
along the ray is

\be t = \pm \int^r \frac{dr}{r}\sqrt{r^2N^2(r) - h^2} + h\phi. \ee In the usual
case (see Fig. 1), in which the angle $\widehat{AOB}$ is obtuse, the ray has two
branches, both taken with the positive sign: an incoming one from $r_{_{A}} $ to $b$
and an outgoing one from $b$ to $r_{_{B}}$. In the acute case $b$ is never
reached and we have just an outgoing ray from $r_{_{A}}$ to $r_{_{B}}$. In both
cases, in going from A to B the longitude increases by $ \phi_{_{B}} -
\phi_{_{A}} = \Phi_{_{AB}}.$ The quantity

\be S(h)= \int^{r_B}_b \frac{dr}{r}\sqrt{r^2N^2(r) - h^2} \pm \int^{_{r_A}}_b
\frac{dr}{r}\sqrt{r^2N^2(r) - h^2} +h\Phi_{_{AB}} \label{eq:lighttime}\ee gives
the phase change, hence the light-time, between the end points, but the
quantity $h$ is still arbitrary. The upper (lower) sign corresponds to the case
in which the angle $\widehat{AOB}$ is obtuse (acute); in the latter case the
two integrals combine in a single one from $r_{_{A}}$ to $r_{_{B}}$, and $b$
disappears as a lower limit. (\ref{eq:lighttime}) is what Fermat's action
functional becomes when its variability is restricted to $h$ and the longitude
constraint is not imposed; it shall be called \emph{reduced action}. At the
true value it satisfies

\be S'(h_{\mathrm{true}}) =0 \label{eq:stationary}, \ee keeping the end points
fixed.

The present work aims at providing the theoretical foundation for the time
delay in all configurations; the sign freedom allows dealing with both cases
at the same time, but applications will be mainly given for a conjunction, with
the $+$. The origin of longitudes is arbitrary. This general approach is relevant, for ex ample, for a spacecraft on an almost parabolic orbit, as in the Solar Probe concept; with a perihelion as low as $4R_\odot$, it can have a strong enhancement of the light-time even in the acute configuration.

\begin{figure}
\includegraphics[width=4in]{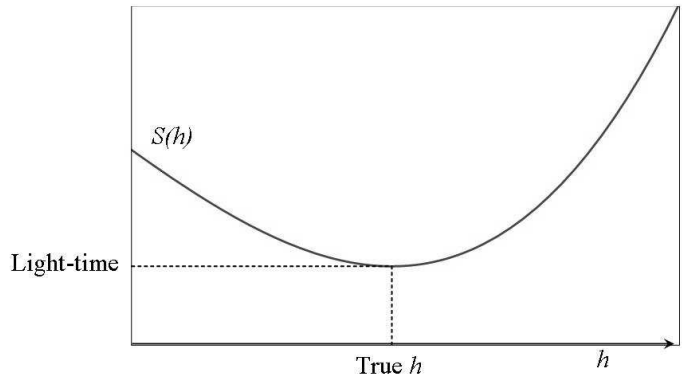}
\caption{The minimum of the reduced action (\ref{eq:lighttime}) is equal to the
light-time at the true value $h_{\mathrm{true}}$.}
\end{figure}

In the derivative $S'(h)$ there are no contributions from the lower limits;
then (\ref{eq:stationary}) provides $h$ as an implicit function of the total total elongation $\Phi_{AB}$:

\be \Phi_{_{AB}} + \int^{r_B}_b\frac{dr}{r}\frac{-h}{\sqrt{(rN(r))^2 - h^2}} \pm
\int^{r_A}_b\frac{dr}{r}\frac{-h}{\sqrt{(rN(r))^2 - h^2}} =0. \label{eq:Phi}\ee Hence
(\ref{eq:lighttime}) reads\footnote{In a slightly inconsistent notation, we often use $h$
to denote both an independent and variable quantity, and the fixed value
$h_{\mathrm{true}}$ determined by the elongation. The context should be sufficient to
clear the ambiguity.}

\be S(h) = \int^{r_B}_b dr N(r) \frac{rN(r)}{\sqrt{(rN(r))^2 - h^2}} \pm \int^{r_A}_b dr
N(r) \frac{rN(r)}{\sqrt{(rN(r))^2 - h^2}}. \label{eq:time}\ee i

Both integrals are convergent (and in the acute case the singularity at $rN(r) = h$ is not even reached). (\ref{eq:lighttime}) suggests the introduction of the
function

\be\label{eq:G(r)} G(r,h)=\int_b^r \frac{dr}{r}\sqrt{(r N(r))^2-h^2}\,, \ee in terms of which
\be S(h)=G(r_B,h) \pm G(r_A,h)+h \Phi_{AB}. \ee
(\ref{eq:Phi}) reads\footnote{The suffix $,_h$ indicates partial derivative.}

\be\label{eq:Phi2} G_h(r_B,h) \pm G_h(r_A,h)+ \Phi_{AB}=0. \ee While in
(\ref{eq:lighttime}) $h$ is an independent parameter, in (\ref{eq:time}) it is
fixed by (\ref{eq:Phi}). 

This expression for $h$ can also be derived directly
from Fermat's Principle, thus providing its significance. Fermat's action
(\ref{eq:action}), expressed as a function of $r$, has the Lagrange functional

\be \mathcal{L}_{_{F}}[\phi(r) ] = N(r) \sqrt{1 + r^2(d\phi/dr)^2},
\label{eq:lagrange}\ee with the (positive) constant of the motion

\be \frac{\partial \mathcal{L}_{_{F}}}{\partial (d\phi/dr)} = \pm \frac{r^2
N(r)}{\sqrt{1 + r^2(d\phi/dr)^2}} \frac{d\phi}{dr}=h .\ee The upper (lower)
holds for the outgoing (incoming) branch. Integrating

\begin{figure}
\includegraphics[width=4in]{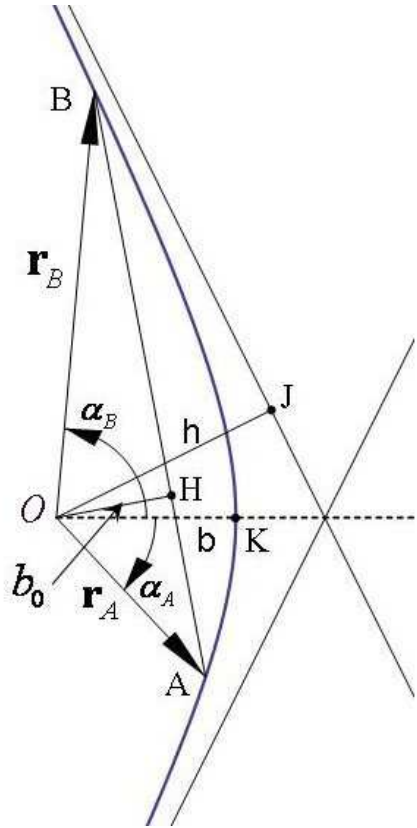}
\caption{Three ways to define the separation of the ray from the origin: the distance
$b_0 = h_0 = OH$ (in this paper often taken as unit of length) of the straight line AB;
the distance $b = OK$ the point of closest approach; the impact parameter $h = bN(b) =
OJ$. }
\end{figure}

\be r\frac{d\phi}{dr} = \pm \frac{h}{\sqrt{r^2N^2(r) - h^2}} = \pm\frac{h}{\sqrt{\rho^2
-h^2}}, \label{eq:h} \ee (\ref{eq:Phi}) is recovered.  Comparison with the Newtonian case (\ref{eq:angmom}) shows that the latter corresponds to the \emph{exact} index of refraction

\be N_{\mathrm{New}}(r) =\sqrt{1 + 2\frac{m}{r}}, \ee corresponding, as
expected, to $\gamma =0 $, $N_1 =1$ and $N_2 = - N_3 =- 1/2$, etc.

\section{A variational argument}

At this point one could proceed as follows:  using power series, solve
(\ref{eq:Phi}) for $h$ in terms of $\Phi_{AB}$, a known quantity. The value of
$h$, inserted into (\ref{eq:time}), provides the required light-time. The
stationary character of the action (\ref{eq:stationary}), however, brings about
a deep and important simplification. This is already tacitly applied in the
usual derivation of the gravitational delay (\ref{eq:standard}). To first
order, the integral of $dt = N(r)d\ell$ in (\ref{eq:isotropic}) reads

$$ t_{_{B}} - t_{_{A}} = \int_{\ell_{_A}}^{\ell_{_B}} d\ell + (\gamma
+1)\int_{\ell_{_A}}^{\ell_{_B}} d\ell\, \frac{m}{r} ; $$ the second integral
can be carried out along the straight path from {\it A} to {\it B}, leading to
the characteristic logarithmic term. In principle, however, the first integral
should take into account the (first order) deflection; we should understand $\int d\ell =\ell_{_{AB}}$ as the Euclidian length of the bent arc between {\it A} and {\it B}. But the length $r_{_{AB}}$ of the straight segment {\it AB} is a minimum in the set of all curves joining {\it A} and {\it B}, so that $\ell_{_{AB}} - r_{_{AB}}$ vanishes to $O(m)$.\footnote{A didactical remark is in order here. This minimum property, crucial to the argument, is often omitted in the usual derivation. See, e. g., \cite{misner73} p. 1107, \cite{ciufolini95} p. 125; in equation (17.59) of \cite{bbpfdv03}, p.  581 the minimum is not mentioned and a factor 4 is missing in the argument of the logarithm.} \emph{Ray bending is irrelevant here.}

In order to exploit the variational nature of the problem it is convenient to
apply power expansions \emph{before} imposing the extremum condition
(\ref{eq:stationary}). We just need the value of the reduced action
(\ref{eq:lighttime}) $ S(h) =\sum_s m^s S_s(h) $ at the value which fulfils
(\ref{eq:stationary}), namely, $ 0 =\sum_s m^s S_s'(h). $ Setting $h = h_0 +
mh_1 + m^2h_2$ and expanding, the solution to second order is obtained iteratively:

\ba S_0'(h_0) &=& 0, \label{eq:h0} \\
h_1S_0''(h_0) &+& S_1'(h_0) = 0,    \label{eq:h1}  \\
h_2S_0''(h_0) &+& \frac{h_1^2}{2}S_0'''(h_0) + h_1 S_1''(h_0) + S_2'(h_0)=0.
\label{eq:h2} \ea In the expression

\ba S(h) &= & S_0(h_0) + m(h_1S_0'(h_0) + S_1(h_0)) + \nonumber \\
& +& m^2 \left (h_2S_0'(h_0) +\frac{h_1^2}{2}S_0''(h_0) + h_1S_1'(h_0) +
S_2(h_0)\right), \label{eq:S2}\ea the effect of the extremum property is
clear: since $S_0'(h_0)=0$, the first order term does not contain $h_1$, and
the second order term does not contain $h_2$; in general, the term in $S(h)$ of
order $m^k$ does not depend on $h_k$.   This important result is reflected in the general approach of \cite{poncinlafitte04}.  Referring to the equation numbering of that paper, their world function $\Omega(x_{_A},x_{_B})$ fulfills the Hamilton-Jacobi equation (30).  In the null case $\Omega=0$, (30) becomes the eikonal equation.  Their Theorem 2 proves that the $n^{\rm th}$-order $\Omega^n$ can be expressed in terms of integrals along the lowest order Minkowskian path.  In our case this \emph{variational Lemma} clarifies the matter and produces considerable simplifications. Using (\ref{eq:h1}), the light-time to second
order reads:

\be S(h) = S_0(h_0) + m S_1(h_0) + m^2\left(\frac{h_1^2}{2}
S_0''(h_0) + h_1S_1'(h_0) + S_2(h_0)\right) \label{eq:S}, \ee where $h_1$ is given
by (\ref{eq:h1}). The delay coefficients (\ref{eq:delayDelta}) read

\be \Delta_1 = S_1(h_0), \quad \Delta_2 = \frac{h_1^2}{2}S_0''(h_0) +
h_1S_1'(h_0) +S_2(h_0). \label{eq:delta}\ee The second-order correction in the
impact parameter $h_2$, given by (\ref{eq:h2}), is needed only at third and
higher orders. For the record, note the third-order contribution to the
light-time:

\be \Delta_3 = h_1h_2S_0'' +\frac{h_1^3}{6}S_0''' + h_2S_1'
+\frac{h_1^2}{2}S_1''  + h_1 S_2' + S_3, \label{eq:S3} \ee where, for
simplicity, the arguments $h_0$ have been understood, and $h_2$ is provided by (\ref{eq:h2}).

\section{Power series}

We now proceed to apply this simple and general Lemma to the light-time. To
lowest order, in (\ref{eq:G(r)}) we use $b = b_0 = h_0$ and $N(r) =1$, so that

\ba S_0(h_0) &=& \sqrt{r_{_{B}}^2 - h_0^2} \pm \sqrt{r_{_{A}}^2 - h_0^2} +
\nonumber \\
&-& h_0\left(\arccos\frac{h_0}{r_{_{B}}} \pm \arccos\frac{h_0}{r_{_{A}}} -
\Phi_{_{_{AB}}}\right). \label{eq:S0} \ea The condition $S_0'(h_0) =0 $ determines
$h_0$ with the trigonometric relation (see Fig. 1)

\be\label{eq:PhiAB} \Phi_{_{_{AB}}} = \arccos\frac{h_0}{r_{_{B}}} \pm
\arccos\frac{h_0}{r_{_{A}}}, \ee or

\be h_0 = \frac{r_{_A} r_{_B}}{r_{_{AB}}} \sin \Phi_{_{AB}}. \ee Therefore

\be S_0(h_0) = \sqrt{r_{_{B}}^2 - h_0^2} \pm \sqrt{r_{_{A}}^2 - h_0^2} =
r_{_{_{AB}}}= \sqrt{r_{_A}^2+r_{_B}^2-2r_{_A} r_{_B} \cos
\Phi_{_{AB}}},\label{eq:S0reduced} \ee is the geometric distance AB. $h_0$ is now
fixed and can be taken equal to unity without loss of generality. Because of
the variational Lemma, at the next order we can retain $h = h_0 =1$; 
(\ref{eq:lighttime}), with $N(r) = 1+ mN_1/r$, reads

\ba S_0(h_0) &+ &m S_1(h_0) = r_{_{AB}} + mN_1
\left[\int_1^{r_{B}}\frac{dr}{\sqrt{r^2 -1}} \pm
\int_1^{r_{A}}\frac{dr}{\sqrt{r^2 -1}}\right] = \nonumber \\
   &=& r_{_{_{AB}}} + mN_1\left[\ln\left(r_{_{B}} +
\sqrt{r_{_{B}}^2 - 1}\right) \pm \ln\left(r_{_{A}} + \sqrt{r_{_{A}}^2 -
1}\right)\right]. \ea In the obtuse case (with the $+$ sign) the logarithm has
the argument

\be \left(r_{_{B}} + \sqrt{r_{_{B}}^2 -1}\right)\left(r_{_{A}} +
\sqrt{r_{_{A}}^2 - 1}\right) =  \frac{r_{_{_A}}+ r_{_{_B}} +
r_{_{_{AB}}}}{r_{_{_A}}+ r_{_{_B}} - r_{_{AB}}}, \label{eq:cross} \ee as easily
checked by cross multiplication using (\ref{eq:S0reduced}); then
the standard expression (\ref{eq:standard}) is properly recovered. (See the
Appendix for the confusion that can arise due to the gauge freedom and the
difference between closest approach and the distance $b_0$). In the acute case,
instead,

$$ t_{_{B}} - t_{_{A}} =  r_{_{_{AB}}} +
mN_1\ln\left(\frac{r_{_{B}} + \sqrt{r_{_{B}}^2 -1}}{r_{_{A}} + \sqrt{r_{_{A}}^2
- 1}}\right). $$

Before proceeding to the next order we need to evaluate $h_1$ with
(\ref{eq:h1}). Differentiating (\ref{eq:S0}) twice we easily get

\be\label{eq:h1N1} h_1\left[\frac{1}{\sqrt{r_{_{B}}^2-1}} \pm
\frac{1}{\sqrt{r_{_{A}}^2-1}}\right] =
N_1\left[\frac{r_{_{B}}}{\sqrt{r_{_{B}}^2-1}} \pm
\frac{r_{_{A}}}{\sqrt{r_{_{A}}^2-1}}\right]. \ee In Sec. 9 the obtuse case in
which $r_{_{A}} \rightarrow \infty$ will be considered; it simply gives

\be h_1 = N_1 r_{_{B}}. \label{eq:Binfinity} \ee Considerable simplification may be
achieved with the aid of the identities:

\be\label{eq:sqrtrBmh} \sqrt{r_{_B}^2-h_0^2}=r_{_B}(r_{_B}-r_{_A} \cos
\Phi_{_{AB}})/r_{_{AB}}; \ee \be\label{eq:sqrtrAmh} \sqrt{r_{_A}^2-h_0^2}= \pm
r_{_A}(r_{_A}-r_{_B} \cos \Phi_{_{AB}})/r_{_{AB}}. \ee In both cases the
expression for $h_1$ becomes

\be\label{eq:h1simple} h_1 =
N_1\big(\frac{r_{_A}+r_{_B}}{r_{_{AB}}}\big)\big(\frac{1 - \cos
\Phi_{_{AB}}}{\sin \Phi_{_{AB}}}\big). \ee It is useful to record the value of
$b_1 = h_1 - N_1$:

\be b_1 \left[\frac{1}{\sqrt{r_{_{B}}^2- 1}} \pm \frac{1}{\sqrt{r_{_{A}}^2-1}}
\right] = N_1\left[\sqrt{\frac{r_{_{B}} -1}{r_{_{B}}+1}} \pm
\sqrt{\frac{r_{_{A}}-1}{r_{_{A}} +1}}\right]. \label{eq:b1}. \ee Enhancement is
at work: in the obtuse case, with the $+$ sign, the elongation comes close to $\pi$ and $h_1$
becomes large, as discussed in the following Section. In the acute case $h_1$
remains of order unity.

At the next order (see (\ref{eq:S2})), we need $G_2(r,h),\ G_1(r,h)$ and its
first derivative with respect to $h$, and $G_0(r,h)$ with its first and second
derivatives (see (\ref{eq:G(r)})).  At order $s$ we need $G_0(r,h)$ with its
first $s$ derivatives.  If these differentiations are carried out \emph{before}
the integration, a technical difficulty arises. $h$ appears both in the lower
limit and in the square root.  In the obtuse case, already at the second order
each of the two contributions diverges; the second derivative of the
integrand, for instance, has a non-integrable term $\propto (r^2-h_0^2)^{-3/2}$; it turns out, however, that this divergence is compensated by the lower limit contribution.  At higher orders the complexity increases. In the acute case the singular point is not within the integration domain and no hindrance arises. This suggests that the integration is best carried out first, leading to a finite result whose differentiation is straightforward.

The hindrance arises because as $m \rightarrow 0$ the singular point at $r=b$
moves. The integration variable

\be u(r)=\frac{r N(r)}{b N(b)} =\frac{\rho}{h}\ee keeps the singularity fixed at
$u=1$ and cures the problem. Note the appearance of Moyer's radial
coordinate

\be \rho =r N(r) = r + mN_1 + m^2\frac{N_2}{r}. \label{eq:rho}\ee  Then
(\ref{eq:G(r)}) reads \be G(r,h) = h \int_1^{u(r)} du \frac{d \ln
r(u)}{du}\sqrt{u^2-1}. \ee $r(u)$, the inverse of $u(r)$, is itself a power
series, so that

\be \frac{d \ln r(u)}{du }=\sum_{s=0} \left(\frac{m}{h}
\right)^s\frac{C_s}{u^{s+1}} = \frac{1}{u} + \frac{m}{h}
\sum_{r=0}\left(\frac{m}{h}\right)^r\frac{C_{r+1}}{u^{r+2}} = \frac{1}{u}
+\frac{m}{h}q(u).\label{eq:lnr} \ee We have split out the main part $1/u$ from
the correction $O(m/h)$. $C_s$ are numbers $O(m^0)$ constructed with the set
\{$N_k$\}:
\be C_0 =1, \;C_1= N_1, \:C_2= N_1^2+ 2N_2, \:C_3= N_1^3+6 N_1 N_2+ 3
N_3, \ldots \label{eq:set}.
\ee
Hence

\be G(r,h)=h\sum_s\left(\frac{m}{h}\right)^s C_s J_s(u)=\sum_s m^s G_s(r,h)\,, \label{eq:G}\ee
where

\be J_s(u)=\int_1^u du \frac{\sqrt{u^2-1}}{u^{s+1}} \ee are elementary functions. Except for constant contributions, their power expansions for large $u$ are odd (even) for $s$ even (odd).  As implied in Eq. (\ref{eq:G}), $h$ is not expanded in the functions $G_s$.

With this general formalism we can draw an interesting conclusion about
enhancement, which corresponds to the limit $(u_{_{_A}},u_{_{_B}}) \gg 1$. When
$u \gg 1$ the functions $J_s(u)$ converge to a finite limit of order unity,
except for $J_0(u)\rightarrow u$ and $J_1(u)\rightarrow \ln u$; hence, when $h$
is fixed, at higher order no enhanced terms arise in $G(r,h)$ and in the
reduced action.  {\it Enhancement occurs only when $h$ itself is expanded and
expressed in in terms of the geometric distances of the end points}, just as it
happens in the case of Newtonian hyperbolic motion.

Using the universal integration variable $u$, the second-order contribution to
the light-time in (\ref{eq:delta}) has been calculated out with the aid of a
computer algebra code. We have

\ba S_2(h_0)& = &\frac{N_1^2}{2}\left(\frac{1}{\sqrt{r_{_B}^2-h_0^2}}
\pm \frac{1}{\sqrt{r_{_A}^2-h_0^2}} \right) +\nonumber\\
& + & \frac{1}{2 h_0}(N_1^2+2 N_2)\left(\arccos \frac{h_0}{r_{_B}} \pm
\arccos\frac{h_0}{r_{_A}}\right)\,. \ea With the help of (\ref{eq:PhiAB}) and
(\ref{eq:sqrtrBmh}-\ref{eq:sqrtrAmh}), this expression reduces to

\be S_2(h_0)=\frac{ N_1^2 r_{_{AB}}^3}{2 r_{_A} r_{_B}(r_{_B}-r_{_A} \cos
\Phi_{_{AB}})(r_{_A}-r_{_B} \cos \Phi_{_{AB}})}+\frac{(N_1^2+2
N_2)}{2h_0}\Phi_{_{AB}}. \ee The last term in (\ref{eq:S2}) requires the
derivatives $S_1'(h_0)$ and $S_0''(h_0)$:

\ba S_1'(h_0)& = &-\frac{ N_1}{h_0}\left(\frac{r_{_B}}{\sqrt{r_{_B}^2-h_0^2}}
\pm \frac{r_{_A}}{\sqrt{r_{_A}^2-h_0^2}}\right) =\nonumber\\
& = & -\frac{ N_1 r_{_{AB}}(r_{_A}+r_{_B})(1- \cos
\Phi_{_{AB}})}{h_0(r_{_B}-r_{_A} \cos \Phi_{_{AB}})(r_{_A}-r_{_B} \cos
\Phi_{_{AB}})}\,; \ea

\ba S_0''(h_0) & =& N_1\left( \frac{1}{\sqrt{r_{_B}^2-h_0^2}}
\pm \frac{1}{\sqrt{r_{_A}^2-h_0^2}} \right) = \nonumber\\
& = &\frac{ N_1 r_{_{AB}}^3}{r_{_A} r_{_B} (r_{_B}-r_{_A} \cos
\Phi_{_{AB}})(r_{_A}-r_{_B} \cos \Phi_{_{AB}})}\,. \ea The last term in
parentheses in (\ref{eq:S2}) is therefore
\be -\frac{1}{2}\frac{S_1'^2(h_0)}{S_0''(h_0)}=-\frac{N_1^2
r_{_{AB}}(r_{_A}+r_{_B})^2 (1-\cos \Phi_{_{AB}})^2}{2 r_{_B} r_{_A} \sin^2
\Phi_{_{AB}}(r_{_B}-r_{_A} \cos \Phi_{_{AB}})(r_{_A}-r_{_B} \cos
\Phi_{_{AB}})}\,. \ee Combining this with the first term in parentheses in
(\ref{eq:S2}), we obtain

\ba  S_2(h_0)- \frac{1}{2} \frac{S_1'^2(h_0)}{S_0''(h_0)}  =-
\frac{ N_1^2 r_{_{AB}}}{r_{_A} r_{_B} (1+\cos \Phi_{_{AB}})}
+\frac{N_1^2+2 N_2}{2h_0} \Phi_{_{AB}} = \hbox to 1.5in{} \nonumber \\
\hbox to .65in {} =-\frac{ N_1^2 r_{_{AB}}}{r_{_A} r_{_B} (1+\cos
\Phi_{_{AB}})}+\frac{ r_{_{AB}}(8-4\beta+8\gamma-3\epsilon)\Phi_{_{AB}}}{4
r_{_B} r_{_A} \sin \Phi_{_{AB}}}\,.\hbox to 1.4in{} \nonumber 
\ea 
The light-time to second order (\ref{eq:S2}) is therefore

\ba t_{_B}-t_{_A} = r_{_{AB}}+m N_1 \ln\big[(r_{_B}+r_{_A} + r_{_{AB}})/(r_{_B}
+ r_{_A}-r_{_{AB}})\big] + \nonumber\\
 m^2 \frac{r_{_{AB}}}{r_{_A} r_{_B}} \left(\frac{N_1^2 +
 2N_2}{2} \frac{\Phi_{_{AB}}}{\sin \Phi_{_{AB}}} - \frac{N_1^2 }{ 1+
\cos \Phi_{_{AB}}} \right); \label{eq:final} \\
t_{_B}-t_{_A} = r_{_{AB}}+m N_1 \ln \big[(r_{_B}+
\sqrt{r_{_B}^2-h_0^2})/(r_{_A}+
\sqrt{r_{_A}^2-h_0^2})\big]  +\nonumber\\
 +m^2 \frac{r_{_{AB}}}{r_{_A} r_{_B}}\left(\frac{N_1^2 +
2N_2}{2}\frac{\Phi_{_{AB}}}{\sin \Phi_{_{AB}}} - \frac{N_1^2 }{ 1+ \cos
\Phi_{_{AB}}}\right), \ea in the obtuse and acute cases, respectively.
Remarkably, $\Delta_2$ has the same expression. This agrees with the result
obtained in \cite{teyssandier08}.

With the same technique, using (\ref{eq:S3}), we have computed also the reduced
action at the third order. For good measure, here is the result:

\ba  S_3(h)&=& \frac{1}{6h^2(r_{_B}^2-h^2)^{3/2}} \left(2N_1^3+6N_1N_2+3N_3)r_{_B}^3\right. +
 \nonumber \\
&-& \left.3 h^2(N_1^3+6 N_1N_2+4 N_3)r_{_B} +6 h^3(N_1N_2+ N_3)\right) +  \nonumber\\
&\pm & \frac{1}{6 h^2 (r_{_A}^2-h^2)^{3/2}}
\left(2(N_1^3+6N_1N_2+3N_3)r_{_A}^3\right. +\nonumber \\
&-& \left. 3 h^2(N_1^3+6 N_1N_2+4 N_3)r_{_A} +6 h^3(N_1N_2+N_3)\right). \ea

\section{Enhancement}

Enhancement occurs in the obtuse case when $r_{_{A}}$ and $r_{_{B}}$ are both
much larger than $b_0 = h_0 =1$, so that (\ref{eq:h1N1}) reduces to

\be h_1\left(\frac{1}{r_{_{A}}}+ \frac{1}{r_{_{B}}}\right)= h_1\frac{2}{R} =2N_1. \ee As hinted in Sec. 2
for the Newtonian case, it is appropriate to formally introduce another infinitesimal
parameter $b_0/R = 1/R$, where $R$ is the harmonic mean of the distances
(\ref{eq:harmonic}). When the ratio $r_{_{A}}/r_{_{B}}$ is $O(R^0)$, as we assume, the
$n$-th order harmonic average $1/r_{_{A}}^n + 1/r_{_{B}}^n $ is $O(1/R^n)$. The
intermediate case $b_0 \approx r_{_{A}} \ll r_{B}$, not discussed here, also shows
enhancement. For instance, it occurs in a nearly parabolic orbit with a small perihelion distance $p_\odot$, as in the case of a solar probe, for which even $p_\odot=4R_\odot$ has been envisaged. The expansion of

\be  h_1 = RN_1 + O(1/R) \ee has only odd terms. One should also note that, as
can be seen from Fig. 1, the angle $\Phi_{_{AB}}$ is fixed by the Euclidean experimental setup and should be considered independent of $m$. In the approximation $h_0=1 \ll R$,

$$\Phi_{_{AB}} = \pi - 2/R + O(1/R^3) $$ is slightly less than $\pi$; the law
of cosines has been used here.

We now proceed to discuss enhancement at the second and third order. It is convenient to
first review the behaviour of the function $G(r,h)$ (\ref{eq:G}) in the limit
$r/h = O(R) \gg 1$. Replacing
$\rho$ with its expression (\ref{eq:moyer}) and expanding, one gets:

\ba && G_0(r,h) = hJ_0(r/h), \quad G_1(r,h) =N_1\big[J_0'(r/h) +
J_1(r/h)\big], \nonumber \\
&& G_2(r,h) =\frac{N_2}{r}J_0'(r/h) +\frac{N_1^2}{2h}\big[J_0''(r/h)+ 2J_1'(r/h)\big] +
\frac{C_2}{h}J_2(r/h).\nonumber \ea Now, when $u \gg 1$
\ba
 J_0(u) =-\frac{\pi}{2}+u + \frac{1}{2u} + \frac{1}{24 u^3} +\ldots, \nonumber\\
J_1(u) = -1 + \ln(2u) +\ldots, \:\: J_2(u) = \frac{\pi}{4}-\frac{1}{u} +\ldots; \nonumber
\ea
setting $u=r/h$,

\ba && G_0(r,h) = r-\frac{\pi h}{2}+\frac{h^2}{2r} + \frac{h^4}{24 r^3}+ \ldots, \: \nonumber\\
&&G_1(r,h) = N_1\left[-\frac{h^2}{4r^2} + \ln\frac{2r}{h} +\ldots\right], \nonumber \\
&& G_2(r,h) = N_1^2\left(\frac{\pi}{4 h}+\frac{h^2}{6r^3} \right)+ N_2\left(\frac{\pi}{2 h}-\frac{h^2}{6r^3}-\frac{1}{r}\right)+\ldots
 \,. \nonumber
\ea We need

\ba && G_{0,hh}(r,h) = \frac{1}{r} + \frac{h^2}{2r^3} \rightarrow \frac{1}{r}, \quad
G_{0,hhh}(r,h) \rightarrow \frac{h}{r^3}, \nonumber \\
&& G_{1,h}(r,h) =-N_1 \left(\frac{1}{h} + \frac{h}{2r^2} \right) \rightarrow
-\frac{N_1}{h}, \nonumber \\
 && G_{1,hh}(r,h) = N_1 \left(-\frac{1}{2r^2} + \frac{1}{h^2}\right)\rightarrow \frac{N_1}{h^2},\nonumber\\
&& G_{2,h}(r,h) = -N_1^2\left(-\frac{\pi}{4 h^2}+\frac{h}{3 r^3}\right)+N_2\left(-\frac{\pi}{2 h^2}-\frac{h}{3 r^3}\right)\nonumber\\
&&\quad\quad\quad\quad \rightarrow -(N_1^2+2 N_2)\frac{\pi}{4 h^2}\,.
\nonumber \ea 

In the expression (\ref{eq:delta}) of $\Delta_2$ the last term is constructed with
$G_2(r,h)$ and is not enhanced. The second term comes from $G_{1,h}(r,h_0) = -N_1$ and,
when summed over the end points, contributes to the light-time with $-2N_1 h_1 = - 2N_1^2
R. $ Lastly, the first term gives $h_1^2/R = N_1^2 R$. Therefore the enhanced part of the
second-order contribution to the light-time is

\be \Delta_{2\mathrm{enh}} = -N_1^2R +O(R^0), \ee in agreement with
(\ref{eq:moyercorrection}). \emph{The second-order terms in the ODP are just the enhanced
ones.}

In a similar way, we get the enhanced third-order terms.  For this we need the enhanced part of $h_2$, to be extracted
from (\ref{eq:h2}); its terms are constructed, respectively, with $G_{0,hh}$, $G_{0,hhh}$,
$G_{1,hh}$ and $G_{2,h}$. Using their asymptotic expressions above one gets the relation

$$ \frac{2}{R}h_2 + \frac{h_1^2}{2}\left(\frac{1}{r_{_{A}}^3} +
\frac{1}{r_{_{B}}^3}\right) + 2h_1 N_1 + (N_1^3 - N_2)\left(\frac{1}{r_{_{A}}^3} +
\frac{1}{r_{_{B}}^3}\right) =0. $$ The third term prevails, and

\be h_2 = -h_1N_1R = - N_1^2 R^2 +O(R), \ee in agreement with the Newtonian case
(which corresponds to $N_1 =1$).

In the expression (\ref{eq:S3}) for $\Delta_3$

\ba
 \Delta_3 = \frac{h_1h_2}{R} + \frac{h_1^3}{6} \left(\frac{1}{r_{_{A}}^3} +
\frac{1}{r_{_{B}}^3}\right) - 2h_2N_1 + h_1^2 N_1\nonumber\\
-\frac{\pi h_1}{2}(N_1^2+2N_2)-3(N_1^3+6N_1N_2+4N_3)(r_{_A}+r_{_B}) 
\ea
the first, third and fourth terms are enhanced, so that finally

\be \Delta_{3\mathrm{enh}} = N_1^3R^2 +O(R). \ee 
Similarly, it turns out that $\Delta_{4\,\mathrm{enh}} \propto N_1^4 R^3 +O(R^2)$.

To summarize, the expansion (\ref{eq:delayDelta}) reads (for the Sun):

\be \frac{\Delta t}{m} =  \Delta_1 + 2\times 10^{-6}\frac{R_\odot}{b_0} \Delta_2
 + 4\times
10^{-12} \left( \frac{R_\odot}{b_0}\right)^2 \Delta_3 + \ldots \ee
In the obtuse case, when $R \gg b_0$, $\Delta_s $ a descending power of
$R/b_0$, beginning with $(R/b_0)^{s-1}$. This is the main enhanced term. It depe
nds
only on the single PPN parameter $N_1$: one could say, enhancement arises due to
 the long-range component $\propto
1/r$ of the index of refraction. $\Delta_1$, typically $\approx 10\,N_1$, is the
 (logarithmically enhanced) term of (\ref{eq:conjunction});

$$ \Delta_2= -N_1^2\left(\frac{R}{b_0} + O(1)\right), \quad \Delta_3=
N_1^3\left[\left(\frac{R}{b_0}\right)^2 + O\left(\frac{R}{b_0}\right)\right] $$
 single out the
main enhanced contribution. For a given $R$, the strongest possible enhancement
occurs when $b_0= R_\odot$; numeriocally

\be  \frac{\Delta t}{m} = 10\,N_1 -2\times 10^{-6} \frac{R}{R_\odot}N_1^2 + 4\times 10^{-12} \left(\frac{R}{R_\odot}\right)^2 + \ldots \,.\ee In a typical configuration, with one station on the Earth, $R_A = 1 AU \ll r_B$, so that $R = 2\,AU = 400\,R\odot$.
The three terms in the expression above are about $20, 3.2\times 10^{-3}, 6.4 \times 10^{-7}$. For a given accuracy in $N_1$ (or $\Delta_1$) this shows how many terms are needed in the expansion in this extreme case.

\section{Deflection}

 In the standard theory, the deflection of the image of a far away source is the acute angle $\delta$ between the asymptotes of the ray. Taking the origin of longitudes on the symmetry axis OK through closest approach (Fig. 4) and using (\ref{eq:Phi}), the longitude of the outgoing asymptote reads (with $\rho = uh$)

\be \phi_{_{\infty}} = \frac{\pi + \delta}{2} = h\int_b^\infty
\frac{dr}{r\sqrt{\rho^2 - h^2}} =  \int_1^\infty du\frac{d\ln
r(u)}{du}\frac{1}{\sqrt{u^2-1}}. \ee Expanding in powers of $m/h$, using 
(\ref{eq:set}) and separating out the main part,

\be \phi_{_{\infty}} = \sum_{s=0} C_s I_s \left(\frac{m}{h}\right)^s =
\frac{\pi}{2} + \frac{m}{h}\sum_{s=1} C_s I_s\left(\frac{m}{h}\right)^s, \ee
where

$$ I_s = \int_1^\infty \frac{du}{u^{s+1}\sqrt{u^2 -1}}  $$ are numerical
constants and $d(\log(r(u))/du$ has been defined in Eq. (\ref{eq:lnr}).  The total deflection is, explicitly
$$ \delta = 2N_1\frac{m}{h} + \pi \frac{N_1^2 +
2N_2}{2}\left(\frac{m}{h}\right)^2 + \frac{4(N_1^3 + 6N_1N_2 +
3N_3)}{3}\left(\frac{m}{h}\right)^3 + \ldots \,.$$ In the more common isotropic
gauge (\ref{eq:rho})

$$ h = bN(b) = b + N_1m + N_2\frac{m^2}{b} + N_3 \frac{m^3}{b^2} + \ldots , $$
and so
\ba \label{eq:totaldeflection}
&& \delta =  2N_1\frac{m }{b}+\frac{\pi(N_1^2+2 N_2)-4N_1^2}{2}\left(\frac{m}{b}\right)^2 + \nonumber \\
& & +  \frac{10N_1^3+ 18 N_1 N_2+ 12N_3-3\pi N_1^3-6 \pi
N_1N_2}{3}\left(\frac{m}{b}\right)^3+ \ldots \,. \ea In terms of the PPN
coefficients and using the expansion of $h$, to second order we have

\be \delta = \frac{2m(\gamma+1)}{h}+\frac{\pi m^2}{4}(8 -4\beta + 3 \epsilon +8\gamma)\,, \ee which
agrees with \cite{fischbach80}; in general relativity, and using the closest
approach $b$,

\be \delta = 4\frac{m}{b} + (15\pi-32) \frac{m^2 }{4 b^2}+\frac{(155-45\pi)m^3}{3 b^3}\,, 
\ee in agreement to second order with \cite{bodenner03}.

\begin{figure}
\includegraphics[width=4in]{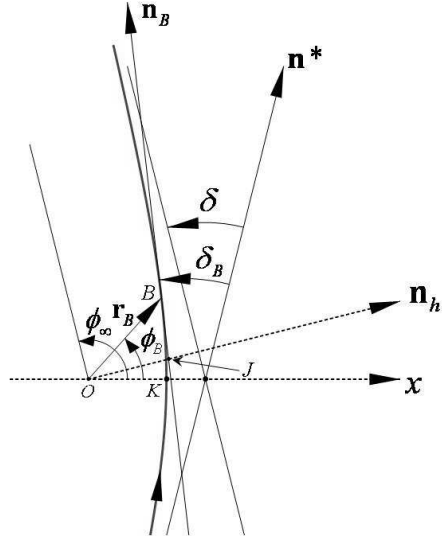}
\caption{Deflection measured from a finite distance. A ray from a far away
source arriving along the direction $\mathbf{n}^\star$ is deflected, and arrives at the observation point {\it B} from a different direction, with unit vector
$\mathbf{n}_{_B}$ (eq. (\ref{eq:tangent})) tangent to the ray. The deflection
angle $\delta_{_B}$ is smaller than the asymptotic deflection $\delta$, the
angle between the asymptotes. The origin of longitudes is taken on the axis {\it OK} through the closest approach, so that $\phi_\infty = \pi/2 + \delta/2$. The
figure also illustrates the meaning of the parameter $h$.  The point {\it J\/}
is the intersection of the tangent through {\it B} with a line through {\it O} perpendicular to the asymptote.  It is easily seen that the distance {\it OJ} = $rh/\sqrt{r^2 N(r)^2-h^2}$ so that at great distance this distance becomes $h$, which therefore is just the \emph{impact parameter}. In the Newtonian dynamical model $h$ (\ref{angmom}) is a constant of the motion, with the same meaning. }
\end{figure}

This standard approach, however, is not adequate for astrometric observations, which are
carried out from a point B at a finite distance $r_{_B}$. In the linear approximation
this problem has been solved in \cite{misner73}, Sec. 40.3; here we give a general
formulation and derive the quadratic term. Referring to Fig. 4, we
need the unit tangent vector $\mathbf{n}(\phi)$ in the counterclockwise direction
(increasing $\phi$) at a generic point $(r\cos\phi, r\sin \phi)$ on the ray (for
simplicity, on the outgoing branch), expressed in terms of the function $r(\phi)$:
\be \mathbf{n}(\phi) =\frac{(r'\cos\phi -r\sin\phi, \,r'\sin\phi
+r\cos\phi)}{\sqrt{r'^2 + r^2}}. \ee From (\ref{eq:h})
$$ r'(\phi) = r \frac{\sqrt{\rho^2 -h^2}}{h}, $$ so that at B, the tangent
vector is
\ba \mathbf{n}_{_{B}} =& \frac{1}{\rho_{_{B}}}\left(\sqrt{\rho_{_{B}}^2
-h^2}\cos \phi_{_{B}} - h\sin \phi_{_{B}}, \, \sqrt{\rho^2_{_{B}}
-h^2}\sin\phi_{_{B}} + h\cos \phi_{_{B}} \right)\nonumber \\
 =&(n_{_{Bx}}, n_{_{By}}).\hbox to 2.7in{} \label{eq:tangent}
\ea 
With
$$ \cos \chi_{_{B}} = h/\rho_{_{B}} = 1/u_{_{B}}, \quad \sin \chi_{_{B}} = \sqrt{\rho^2_{_{B}}
-h^2}\big/\rho_{_{B}} = \sqrt{1- 1/u_{_{B}}^2}, $$ it is convenient to
introduce the quantity $\chi_{_{B}}$, a function on the ray; in the limit
$m\rightarrow 0$, since $\rho \rightarrow r $ and $h \rightarrow 1$, it reduces
to $\alpha_{_{B}}$ (Fig. 1). Then
\be \mathbf{n}_{_{B}}=(\sin(\chi_{_{B}} - \phi_{_{B}}),\, \cos(\chi_{_{B}}
-\phi_{_{B}})). \ee The deflection $\delta_{_{B}}$ is provided by the vector
product
$$\vert \mathbf{n}^\star \times \mathbf{n}_{_{B}}\vert = \sin \delta_{_{B}},
$$ where $\mathbf{n}^\star =(\sin(\delta/2),\cos(\delta/2))$
is a unit vector along the asymptote of the incoming ray. Hence we obtain the exact expression
\be \delta_{_{B}} = \phi_{_{B}} - \chi_{_B} +\frac{\delta}{2}.
\label{eq:deflectionBexact}
\ee 
Two effects contribute in (\ref{eq:deflectionBexact}): a local term $\chi_{_{B}}$ due
to the change in the tangent, and a change in the orientation of the outgoing
asymptote relative to OA. In the case of GAIA and other space astrometric
projects no images can be obtained near the Sun, so that $r_{_{B}} = 1$ AU
$\approx h$ and there is little enhancement. The data analysis will be truly
global, with subtle statistics. The expected angular measurements error
$\approx 5 \times 10^{-11}$ is quite below the first-order deflection $\approx
4 \times 10^{-8}$ and much larger than the second-order term $\approx
10^{-16}$; but the fractional difference between $\delta$ and $\delta_{_{B}}$
is not small.  With our powerful formalism the derivation of the second-order approximation to $\delta_B$ is straightforward. 

Two limits are noteworthy. When $m \rightarrow 0$, $\phi_{_B} $ tends to $\alpha_{_{B}}$ and, of course, there is no
deflection. To recover the standard expression when B goes to infinity, note
that, using \ref{eq:lnr}),
$$ \phi_{_{B}} = \int_1^{u_B}du \frac{d\ln r(u)}{du}\frac{1}{\sqrt{u^2-1}} =
\chi_{_{B}}+ \frac{m}{h}\int_1^{u_B} du \frac{q(u)}{\sqrt{u^2 -1}};$$ therefore to second order
\be
\phi_{_B} =\big(1+\frac{m^2 C_2}{2h^2}\big) \chi_{_B}+\frac{mC_1\sqrt{u_{_B}^2-1}}{h u_{_B}}+\frac{m^2C_2 \sqrt{u_{_B}^2-1}}{2 h^2 u_{_B}^2}\,.
\ee
Thus, the deflection reads
\be\label{eq:deflectionBB}
\delta_{_B}=\frac{\delta}{2}+\big(\frac{m^2 C_2}{2h^2}\big)\chi_{_B}+\frac{mC_1\sqrt{u_{_B}^2-1}}{h u_{_B}}+\frac{m^2C_2 \sqrt{u_{_B}^2-1}}{2 h^2 u_{_B}^2}\,.
\ee
In the limit $u_{_B}\rightarrow \infty$, this agrees with Eq. (\ref{eq:totaldeflection}).  In terms of $r_{_B}$ and $h_0$, this is
\ba
\delta_B=\frac{\delta}{2}&+&\sqrt{r_{_B}^2-h_0^2}\frac{m}{h_0 r_{_B}}\big(C_1+\frac{m C_2}{2 r_{_B}}\big)\nonumber\\
&+&\frac{m^2 C_2 \chi_{_B}}{2 h_0^2}+\frac{m^2 C_1}{\sqrt{r_{_B}^2-h_0^2}}\big(\frac{N_1 h_0}{r_{_B}^2}-\frac{h_1 r_{_B}}{h_0^2}\big)\,.
\ea
For $u_{_B}$ finite, this result agrees in first order with \cite{misner73}, Eq. 40.11.

\section{Conclusion}

With the implementation of optical lasers in deep space, experimental gravity
will undergo a big leap. The planned mission ASTROD (\cite{ni07},
\cite{ni06} and other papers) will consist in a fleet of three drag-free
spacecraft in a triangular configuration with semi-major axes of about 1 AU.
Although no detailed error analysis is available, ranging accuracies of $3
\times 10^{-3}$ cm or better are expected; with closest approach less than 1
AU, this error is comparable with, or smaller than, the second-order
gravitational delay.

Optical interferometry in space will make huge improvements in phase
measurements possible. The GAME (Gamma Astrometric Measurements Experiment)
project (see \cite{gai08}) consists in a Fizeau interferometer in the focal plane
of a space telescope to measure the angular separation of stars in a narrow
field of view near the Sun. The expected accuracy in $\gamma$ of $10^{-7}$ will
require second-order corrections in the gravitational delay.

LISA -- a planned mission for low frequency gravitational wave detection
(\cite{folkner98}, \cite{esa00} and many other papers, in particular \cite{moore01}) --
will fly three drag-free spacecraft orbiting at 1 AU at the vertices of an equilateral
triangle with sides $L = 5\times 10^{11}$ cm; this fleet will rotate around its
centre with the period of a year. Three optical interferometers with baseline $L$ will
operate simultaneously, with an expected sensitivity $\sigma_{_{L}}/L \approx 10^{-21}$
or better. The change in light-time difference between two arms due to the solar
gravitational delay has the period of six months, in a frequency band overwhelmed by the
acceleration noise, but it is interesting to evaluate the effect. For two vertices A and
B, $r_{_{B}} - r_{_{A}} = \delta r \approx L \ll (r_{_{A}}, r_{_{B}}) = 1$ AU. In the
(now generic) acute case the reduced action (\ref{eq:lighttime}) (with the $-$ sign!) is
of order

$$ r_{_{AB} } +mN_1\frac{\delta r}{\sqrt{r_{_{A}}^2 -h_0^2}} \approx 5\times 10^{11} \mathrm{cm }
+ 10^4 \mathrm{cm}. $$ With the approximation $\delta r \ll 1$ AU the action reads

\ba S(h)& =& h\Phi_{_{AB}} + \frac{\delta r}{r_{_{A}}}\sqrt{r_{_{A}}^2N^2(r_{_{A}}^2)
-h^2} = \nonumber \\
&=&  h \Phi_{_{AB}} + \frac{\delta r}{r_{_{A}}}\left[\sqrt{r_{_{A}}^2 - h^2} +m
\frac{N_1
r_{_{A}}}{\sqrt{r_{_{A}}^2 - h^2}}\right.  \nonumber \\
&+& \left. \frac{m^2}{2}\left(\frac{N_1^2 + 2N_2}{\sqrt{r_{_{A}}^2 - h^2}} -
\frac{N_1^2 r_{_{A}}^2}{(r_{_{A}}^2 - h^2)^{3/2}}\right)\right] , \ea an
expression which can be used directly to obtain all relevant quantities. For an
estimate, however, it suffices to remark that in the above $m$-expansion each
term is smaller than the previous one by $O(m/r_{_{A}}) = 10^{-8}$; hence for
LISA the first-, second- and third-order corrections to the light-time are,
respectively, of order $10^4$ cm, $10^{-4}$ cm and $10^{-12}$ cm, corresponding
gravitational wave signals of order

$$ 2\times 10^{-8}, \quad 2\times 10^{-16}, \quad 2\times  10^{-24}. $$ We did not
investigate the consequences of this large, but low-frequency signal on the
performance of the  instrument.

The puzzle of the ODP expression for the gravitational delay has been understood. It must be considered in the framework of an expansion in powers of $m/b$; of all second-order terms so arising, in a close conjunction some are enhanced.  They can be rigorously singled out with a further expansion in diminishing powers of $R/b_0$; those that appear in the ODP are just those of order $m(m/bi_0)(R/b_0)$.  With the powerful tool of geometrical optics, we have provided a procedure to extend the calculation to higher order and have obtained the full correct second-order term of the delay.
  
A methodological reflection is a fit conclusion. The evaluation of the gravitational
delay, a conceptually simple and straightforward problem, faces subtle mathematical
difficulties and a great algebraic complexity. Our approach is based upon two unusual
mathematical levels of description: light propagation with the eikonal theory, rather
than null geodesics, and asymptotic power series, an abstract mathematical tool. The latter, in which ordinary functions
are set aside and an abstract mathematical tool is employed, seemingly runs against
physical intuition. As shown, both are essential to directly attain, and take advantage
of, the crucial features of the problem: the light-time as the minimum of Fermat's
action, and a safe and automatic procedure to select and estimate different terms. This
is another example of the tenet that \emph{every physical problem has an appropriate,
often not intuitive, level of mathematical description}, and severe penalties are in
store for its neglect.

\section*{Appendix}

The radial gauge freedom and the difference between closest approach and $b_0$
can cause some confusion. For example, the textbook \cite{weinberg72} presents
(eq. (8.7.4)) the light-time between closest approach and a generic point; it is
expressed in Schwarzschild's gauge $\bar{r}$ and reads

$$ t(\bar{r},\bar{b}) = \sqrt{\bar{r}^2 -\bar{b}^2}+ (1+\gamma)m\ln\frac{\bar{r}
+\sqrt{\bar{r}^2 -\bar{b}^2}}{\bar{b}} +m\sqrt{\frac{\bar{r} -\bar{ b}}{
\bar{r} +\bar{b}}}, $$ quite different than (\ref{eq:standard}). In a real case
two such terms are needed, one for each branch. But, contrary to what stated in
the textbook, the sum of the two square roots (first term) \emph{is not the
distance AB}. The isotropic gauge and the distance $b_0$, not the closest
approach, should be used. First, setting (\ref{eq:gauge}) $\bar{r} = r
+\gamma m$, the formula reads, to $O(m)$,

$$t(r,b) = \sqrt{r^2 - b^2}+(1+\gamma)m\left(\ln\frac{r +\sqrt{r^2 -b^2}}{b} +
\sqrt{\frac{r-b}{r+b}}\right). $$ Both formulas are useless, however, because the
closest approach $b= b_0 + mb_1 = 1 + mb_1$ is not known beforehand. The ray
must be anchored to two known points and, with $b_1$, is determined by the
unknown $\gamma$ with (\ref{eq:b1}). Since

$$ \sqrt{r^2- b^2} = \sqrt{r^2 -1} - m\frac{b_1}{\sqrt{r^2 -1}} , $$

\ba \sqrt{r_{_{A}}^2- b^2} +\sqrt{r_{_{B}}^2- b^2} &=&
 r_{_{AB}} -mb_1\left(\frac{1}{\sqrt{r^2_{_{A}}-1}} + \frac{1}{\sqrt{r^2_{_B}-1}}\right) =\nonumber \\
&=& r_{_{AB}} - m(1+\gamma)\left(\sqrt{\frac{r_{_{A}}-1}{r_{_{A}}+1}} +
\sqrt{\frac{r_{_B}-1}{r_{_B} +1}}\right), \nonumber \ea and the standard formula is
recovered.

\section*{List of symbols}

\begin{tabular}{ll}

A & event or point where the photon starts \\

$A(r)$  & metric coefficient \\

B & event or point where the photon is detected \\

$B(r)$ & metric coefficient \\

$b$ & closest approach in isotropic variable \\

$C(r)$ & metric coefficient \\

$h$ & closest approach in Moyers's variable \\

$b_0$ & Euclidian approximation of the same \\

$\ell$ & Euclidian arc length \\

$m$ & gravitational radius \\

$N(r) = \sqrt{\frac{B(r)}{A(r)}}$ & index of refraction \\

ODP & Orbit Determination Program  \\

$p_\odot$  & perihelion distance \\

$R = \frac{2r_{_{A}}r_{_{B}}}{r_{_{A}} + r_{_{B}}}$ & harmonic mean of the distances\\

$r$ & isotropic radial coordinate \\

$R_\odot$ & radius of the Sun \\

$\mathbf{r}(\ell)$ & photon trajectory \\

$p_\odot$  & perihelion distance \\

$S$ & Fermat's action \\

$\mathfrak{S}(x^\mu)$ & eikonal function \\

 $t$ & time in the rest frame of the mass \\

$t_{_{A}}$ & starting time of photon \\

$t_{_{B}}$ & arrival time of photon \\

$\gamma$  & relativistic PPN coefficient\\

$\Delta t$ & gravitational delay \\

$\Delta_s $ & expansion coefficients of delay (\ref{eq:delayDelta})\\

$\lambda$ & undefined parameter along the light path \\

$\rho = rN(r)$ & Moyer's radial coordinate \\

$\phi$ & longitude \\

$Phi$ & longitude \\

\end{tabular}


\begin{thebibliography}{99}

\bibitem{bel94} Bel, Ll and Martin J 1994 Fermat's Principle in General Relativity \textit{Gen Rel Grav} \textbf{26} 567-585

\bibitem{bbnali07} Bertotti B, Ashby N and Iess L 2008 The effect of the motion
of the Sun on the light-time in interplanetary relativity experiments
\textit{Class. Quantum Grav. } \textbf{25} 045013 (11 pp)

\bibitem{bbgcli93} Bertotti B, Comoretto G and Iess L 1993 Doppler tracking of spacecraft
with multi-frequency links  \textit{Astron. Astrophys.} \textbf{269} 608-616

\bibitem{bbgg92} Bertotti B and Giampieri G 1992 Relativistic effects for Doppler
measurements near solar conjunction \textit{Class. Quantum Grav. } \textbf{9} 777-793

\bibitem{bbpfdv03} Bertotti B, Farinella P and Vokrouhlick\'{y} D 2003
\textit{Physics of the solar system} (Dordrecht: Kluwer)

\bibitem{bblipt03} Bertotti B Iess L and Tortora P 2003 A test of general
relativity using radio links with the Cassini spacecraft \textit{Nature} \textbf{425}
374-376

\bibitem{bodenner03} Bodenner J and Will C M 2003 Deflection of light to
second order: a tool for illustrating principles of general relativity \textit{Am. J.
Phys.}

\bibitem{born64} Born M and Wolf E 1964 \emph{Principles of Optics} Pergamon
Press

\bibitem{ciufolini95} Ciufolini I and Wheeler J A 1995 \emph{Gravitation and
inertia}. Princeton: Princeton University Press 1995

\bibitem{epstein80} Epstein R and Shapiro I I 1980 Post-post Newtonian deflection of light by the Sun.
\textit{Phys. Rev.} \textbf{D 22}, 2947-2949

\bibitem{erdely56} Erdély A 1956 \emph{Asymptotic expansions}. New York: Dover
Publications

\bibitem{esa00} European Space Agency 2000 \emph{LISA. Laser interferometer space antenna.
System and technology study report} ESA-SCI(2000)11

\bibitem{fischbach80} Fischbach E and Freeman B S 1980 Second-order contribution
to the gravitational deflection of light. \textit{Phys. Rev.}\textbf{D 22},
2950-2952

\bibitem{folkner98} Folkner W M (editor) 1998 Laser interferometer space antenna. Second international LISA
Symposium. AIP Conference Proceedings 456

\bibitem{gai08} Gai M, Lattanzi M G, Ligori S and Vecchiato A 2008 GAME: Gamma Astrometric
Measurement Experiment {\it Proc of SPIE} \textbf{7010} 701027 11 pages

\bibitem{hinch91} Hinch E J 1991 \emph{Perturbation methods}. Cambridge University Press

\bibitem{kopeikin07} Kopeikin S M, Polnarev A G, Sch\"{a}fer G and Vlasov I Yu 2007
Gravimagnetic effect of the barycentric motion of the Sun and determination of the
post-Newtonian parameter $\gamma$ in the Cassini experiment  \textit{Phys. Lett. A}, \textbf{367} 276-280

\bibitem{poncinlafitte04} C Le Poncin-Lafitte, Linet B and Tyssandier P 2004 World function and time transfer: general post-Minkowskian expansions \textit{Class. Quantum Gravity}, \textbf{21} 4463-4483

\bibitem{misner73} Misner C W, Thorne K S and Wheeler J A 1973 \emph{Gravitation}.
San Francisco: W. H. Freeman

\bibitem{moore01} Moore A T and Hellings R W 2001 Angular resolution of space-based
gravitational wave detectors \textit{Phys. Rev.} \textbf{D 65} 062001

\bibitem{moyer00} Moyer T D 2000 \emph{Formulation for observed and computed
values of Deep Space Network data types for navigation}

\bibitem{ni06} Ni W-T, Bao Y, Dittus H \textit{et al} 2006
\emph{ASTROD I: Mission concept and Venus flyby} \textit{Acta Astronautica}, \textbf{59}
598-607

\bibitem{ni07}Ni W-T 2007 ASTROD (Astrodynamical Space Test of Relativity using Optical
Devices) and ASTROD I \textit{Nucl. Phys. Proc. Suppl.} \textbf{166} 153-158

\bibitem{reasenberg79} Reasenberg R D, Shapiro I I, MacNeil P E 1979 \textit{et al}
Viking relativity experiment: verification of signal retardation by solar gravity
\textit{Astrophys. J.} \textbf{234} L 219-221

\bibitem{richter82} Richter G W and Matzner R A 1982 Second-order contributions to
the gravitational deflection of light in the parametrized post-Newtonian formalism
\textit{Phys. Rev.} \textbf{D 26}, 1219-1224

\bibitem{richter83} Richter G W and Matzner R A Second-order contributions to
relativistic time delay in the parametrized post-Newtonian formalism \textit{Phys. Rev.}
\textbf{D 28} 3007

\bibitem{shapiro64} Shapiro I I. Fourth test of general relativity.
\textit{Phys. Rev. Lett.} \textbf{13}, 789-791, 1964

\bibitem{soffel03} Soffel M H, Klioner S A, Petit G \emph{et al} 2003
The IAU 2000 resolutions for astrometry, celestial mechanics, and metrology in the
relativistic framework: explanatory supplement. \textit{Astr. J.} \textbf{126}, 2687-2706

\bibitem{teyssandier08} Teyssandier P, Le Poncin-Lafitte C (2008) General post-Minkowskian expansion of time transfer functions {\it Class. Quant. Grav.} \textbf{25} 145020 (10pp)

\bibitem{weinberg72} Weinberg S 1972 \textit{Gravitation and cosmology:
principles and applications of the general theory of relativity}. New York: J. Wiley, 1972

\bibitem{will93} Will C M 1993 \textit{Theory and experiment in gravitational
physics} Cambridge: Cambridge University Press

\end{thebibliography}
\end{document}